\documentclass{elsart}
\usepackage{graphicx}

\begin{document}

\begin{frontmatter}

\title{Coupled-channel study of $\gamma p \rightarrow K^+\Lambda$}

\author[Pitt,NTU]{Wen-Tai Chiang},
\author[Pitt]{F. Tabakin}
\address[Pitt]{Department of Physics and Astronomy, University of
               Pittsburgh, PA 15260, USA}
\address[NTU]{Department of Physics, National Taiwan University,
              Taipei 10617, Taiwan}
\author[ANL]{T.-S. H. Lee},
\author[Saclay]{B. Saghai}
\address[ANL]{Physics Division, Argonne National Laboratory, Argonne,
              IL 60439, USA}
\address[Saclay]{Service de Physique Nucl\'{e}aire, DAPNIA-DSM,
                 CEA/Saclay, F-91191 Gif-sur-Yvette, France}

\begin{abstract}
A coupled-channel (CC) approach has been developed to investigate kaon
photoproduction on the nucleon. In addition to direct $K^+ \Lambda$
production, our CC approach accounts for strangeness production including
$K^+ \Lambda$ final state interactions with both $\pi^0 p$ and $\pi^+ n$
intermediate states. Calculations for the $\gamma p \rightarrow K^+
\Lambda$ reaction have been performed, and compared with the recent data
from SAPHIR, with emphasis on the CC effects. We show that the CC effects
are significant at the level of inducing 20\% changes on total cross
sections; thereby, demonstrating the need to include $\pi N$ channels to
correctly describe the $\gamma p \rightarrow K^+ \Lambda$ reaction.
\end{abstract}


\end{frontmatter}

\section{Introduction} \label{sec:Intro}
A major issue in strong interaction physics is to understand baryon
spectroscopy. Meson-baryon scattering has been the predominant reaction
used to study the properties of nucleon resonances ($N^*$). An appealing
alternative is to use electromagnetic probes, e.g., $\gamma N \rightarrow
N^* \rightarrow \pi N$. In this photon-induced process, the relative
weakness of the electromagnetic interaction allows one to use first-order
descriptions of the incident channel, thus making possible more reliable
extraction of $N^*$ information from data. With the recent development of
new facilities such as Jefferson Lab, ELSA, GRAAL, MAMI, and SPring-8, it
is now possible to obtain accurate data for meson electromagnetic
production, including spin-dependent observables.

Among meson photoproduction processes, pion photoproduction is by far the
most studied theoretically and experimentally. However, increased effort
has also been devoted in recent years to investigate kaon photoproduction.
Such studies are motivated by several considerations: (1) the production of
strangeness associated with kaons ($K$) and hyperons ($Y$) allows one to
study the role played by $s$ quarks versus $u$ and $d$ quarks; (2) higher
mass resonances can be better studied by investigating the $N^* \rightarrow
K \Lambda$ and $N^* \rightarrow K \Sigma$ decays; (3) the so-called
``missing resonances''~\cite{Capstick:2000,Bijker:2000} predicted by quark
models, but not observed in $\pi N$ scattering, might be found in kaon
photoproduction since they may couple strongly to $K \Lambda$ and $K
\Sigma$ channels.

At photon laboratory energies from the kaon photoproduction threshold of
$\sqrt{s}=$ 1.61 GeV to about 2.5 GeV, the isobar model is most widely used
for extracting $N^*$ parameters from the kaon photoproduction
data~\cite{Adelseck:1986fb,Adelseck:1990ch,Williams:1992tp,%
Mart:1995wu,David:1996pi,Hsiao:2000ez}. This model is based on an effective
Lagrangian approach in which a number of tree diagrams are evaluated with
coupling constants partly fixed from independent hadronic and
electromagnetic data. Although the isobar models describe the existing kaon
photoproduction data fairly well, multi-step or coupled-channel (CC)
effects due to intermediate $\pi N$ states are ignored. The sequence
$\gamma N \rightarrow \pi N \rightarrow K Y$ in kaon photoproduction can be
substantial, since $\gamma N \rightarrow \pi N$ amplitudes are much greater
than the direct $\gamma N \rightarrow K Y$ amplitudes. If this is indeed
the case, the $N^*$ parameters obtained from one-step isobar models can not
be directly compared with the predictions from existing hadron models. The
importance of the final-state interaction(FSI) in interpreting the $\Delta$
resonant parameters extracted from $\gamma N \rightarrow \pi N$ data has
been demonstrated in Ref.~\cite{Sato:1996gk}. In this work,  we investigate
the same problem concerning the role of CC final state interactions in kaon
photoproduction.

Among existing studies of kaon photoproduction, coupled-channel effects
have been investigated using two approaches. Kaiser {\it et
al.}~\cite{Kaiser:1997js} applied a coupled-channel approach with chiral
SU(3) dynamics to investigate pion- and photon-induced meson production
near the $KY$ threshold. Although their recent
results~\cite{CaroRamon:1999jf} include p-wave multipoles, and thus
reproduce data slightly above the threshold region, their chiral SU(3)
dynamics model can not provide the higher partial waves that are important
in describing the data at higher energies. A similar approach has also been
taken in Ref.~\cite{Oset:1999ri}.

The second CC approach was developed by Feuster and
Mosel~\cite{Feuster:1998cj}. They used a $K$-matrix method to investigate
photon- and meson-induced reactions, including $\gamma p \rightarrow K^+
\Lambda$. In the $K$-matrix approach, all intermediate states are put
on-shell and hence the important off-shell dynamical effects can not be
accounted for explicitly. The advantage of their approach is its numerical
simplicity in handling a large number of coupled channels. However, the
extracted $N^*$ parameters may suffer from the difficulties in interpreting
them in terms of extant hadron models, such as the constituent quark model.

In this paper, we present a dynamical CC model in which the meson-baryon
off-shell interactions are defined in terms of effective Lagrangians. This
is achieved by a direct extension of the existing dynamical
models~\cite{Sato:1996gk,Yang:1985yr,Tanabe:1985rz,Nozawa:1990pu} for pion
photoproduction to include $KY$ channels. We follow the approach developed
by Sato and Lee~\cite{Sato:1996gk}. In this first attempt, we do however
need to make several simplifications since it is a rather complex task to
deal simultaneously with the meson-baryon and photon-baryon intertwined CC
problems. First,  we adopt an existing isobar model developed earlier by
Williams, Ji and Cotanch (WJC)~\cite{Williams:1992tp} as initial input for
the direct $\gamma p \rightarrow K^+\Lambda$ process. This fixes the number
of $N^*$ to be considered and the leading tree-diagrams associated with
strange particles. Second, we use the $\gamma N \rightarrow \pi N$ and $\pi
N \rightarrow \pi N$ partial-wave amplitudes from the VPI partial-wave
analysis~\cite{Arndt:1995bj,Arndt:1996ak} to define the amplitudes
associated with the $\pi N$ channel. This drastically reduces the amount of
data we have to confront in the coupled-channel approach. However, the
strong interaction matrix elements of $K Y \rightarrow K Y$ and $\pi N
\rightarrow K Y $ transition operators are derived rigorously from
effective Lagrangians using the unitary transformation method of
Ref.~\cite{Sato:1996gk}. This derivation marks our major differences with
Kaiser {\it et al.}~\cite{Kaiser:1997js} since we include all relevant
higher partial waves and our approach is applicable at all energies. We
solve the resulting CC equations with numerical precision to account for
the meson-baryon off-shell interactions. The dynamical content of our
approach is clearly very different from the $K$-matrix model of Feuster and
Mosel~\cite{Feuster:1998cj}.

Our CC approach is defined by the starting Lagrangian and by the several
approximations such as the Sato-Lee unitary transformation and a
three-dimensional reduction of the Bethe-Salpeter equation. Such
approximations have been pursued successfully by numerous authors in order
to solve difficult strong interaction problems starting from relativistic
field theory. Lagrangian-based dynamics differs considerably from models
motivated by the $S$- or $K$-matrix approach based on tree-diagrams, in
which the dynamics is partly defined by postulating crossing
symmetry~\cite{CrossSym}. While S-matrix plus tree-diagram approaches can
be a very useful working tool, no solid proof exists that crossing symmetry
can be derived ``exactly'' from relativistic quantum field theory. The best
effort is a perturbative derivation in a simple model, such as presented in
Bjorken and Drell. Thus, our approach is not expected to respect crossing
symmetry, although the driving terms of the employed coupled-channel
scattering equations can be made to have crossing symmetry. Compared with
the previous models based on tree-diagrams supplemented by crossing
symmetry~\cite{Williams:1992tp,David:1996pi,Mizutani:1997sd}, the
coupled-channel approach can satisfy ``dynamically''  the multi-channel
unitarity condition. The CC approach offers the advantage of including
strong dynamics for both kaon photoproduction and kaon radiative decays,
while also incorporating cusp structure due to channel coupling, which is
not described by tree-diagram based theories. Therefore,  there are
advantages to the CC approach that we believe out-weigh the desire for a
crossing symmetric theory.

\section{Coupled-channel approach} \label{sec:CC}
A coupled-channel framework for studying kaon photoproduction can be
obtained straightforwardly by generalizing the dynamical approach developed
by Sato and Lee~\cite{Sato:1996gk} in their investigation of $\pi N$
scattering and pion photoproduction. By adding the $KY$ channel and
appropriate $N^*$ states to their formalism, one can show that the
collision matrix, $A_{KY,\gamma N}$, of the $\gamma N\rightarrow KY$
reaction can be written in operator form as
\begin{eqnarray} \label{eq:TgK}
  A_{KY,\gamma N} = R_{KY,\gamma N} + a_{KY,\gamma N}\,.
\end{eqnarray}
Here $R_{KY,\gamma N}$ denotes the resonant part (to be specified later),
and the nonresonant part $a_{KY,\gamma N}$ is defined by
\begin{eqnarray} \label{eq:ab"}
  a_{KY,\gamma N} &=& b_{KY,\gamma N} \nonumber \\
  & & + \sum_{K'Y'}
  t_{KY,K'Y'}\;G_{K'Y'}^{(+)}\;b_{K'Y',\gamma N}
  + \sum_{\pi N}
  t_{KY,\,\pi N}\;
  G_{\pi N}^{(+)}\;b_{\pi N,\gamma N}\,,
\end{eqnarray}
where $G_\alpha^{(+)}$ is the meson-baryon propagator for channel $\alpha.$
Here $b_{KY,\gamma N}$ and $b_{\pi N,\gamma N}$ are the nonresonant
photoproduction operators for $KY$ and $\pi N$ respectively. The scattering
operators $t_{KY,K'Y'}$ and $t_{KY,\pi N}$ describe the nonresonant parts
of the final $KY\rightarrow K'Y'$ and $ \pi N \rightarrow KY$ transitions,
respectively. Obviously, the third term of Eq.~(\ref{eq:ab"}) contains the
coupled-channel effects due to the intermediate $\pi N$ channel.

To see the dynamical feature of our CC approach, we now combine the
resonant term $R_{KY,\gamma N}$ with the nonresonant operator $b_{KY,\gamma
N}$ and define $B_{KY,\gamma N} \equiv R_{KY,\gamma N} + b_{KY,\gamma N}$.
Eqs.~(\ref{eq:TgK})-(\ref{eq:ab"}) then become
\begin{eqnarray} \label{eq:AB"}
  A_{KY,\gamma N} &=&B_{KY,\gamma N} \nonumber \\
  & & + \sum_{K'Y'} t_{KY,\,K'Y'}\;G_{K'Y'}^{(+)}\;b_{K'Y',\gamma N}
  + \sum_{\pi N}
  t_{KY,\,\pi N}\;G_{\pi N}^{(+)}\;
  b_{\pi N,\gamma N}\,.
\end{eqnarray}
The nonresonant meson-baryon transition operators $t_{KY,\,KY}$ and
$t_{KY,\,\pi N}$ are defined by the following CC equations:
\begin{eqnarray} \label{eq:LS1}
  t_{KY_f,\,KY_i} &=& v_{KY_f,\,KY_i} \nonumber \\
  & &+ \sum_{KY} v_{KY_f,\,KY}\;
    G_{KY}^{(+)}\; t_{KY,\,KY_i}
  + \sum_{\pi N} v_{KY_f,\,\pi N}\;
    G_{\pi N}^{(+)}\;t_{\pi N,\,KY_i}\,,
\end{eqnarray}
\begin{eqnarray} \label{eq:LS2}
  t_{KY_f,\;\pi N_i} &=& v_{KY_f,\;\pi N_i} \nonumber \\
  & &+ \sum_{KY} v_{KY_f,\,KY}\;
    G_{KY}^{(+)}\; t_{KY,\;\pi N_i}
  + \sum_{\pi N}\ v_{KY_f,\,\pi N}\;
    G_{\pi N}^{(+)}\;t_{\pi N,\;\pi N_i}\,.
\end{eqnarray}
The above equations define the off-shell scattering amplitudes. Clearly,
the amplitudes $t_{KY,K'Y'}$ and $t_{KY,\pi N}$, which are needed to
evaluate the second and third terms of Eq.~(\ref{eq:AB"}), can be obtained
from solving Eqs.~(\ref{eq:LS1})-(\ref{eq:LS2}) if the potentials
$v_{KY,K'Y'}$ and $v_{KY,\pi N}$ and the nonresonant $\pi N$ amplitude
$t_{\pi N,\pi N}$ can be calculated from a model. We now note that if the
last two terms in the right-hand-side of Eq.~(\ref{eq:AB"}) are neglected,
our CC model reduces to those previously developed isobar models for which
$A_{KY,\gamma N} = B_{KY,\gamma N}$. In that limit, the resonances are then
included in the $R$ term.

The nonresonant meson-baryon $t$-matrix defined by
Eqs.~(\ref{eq:LS1})-(\ref{eq:LS2}) is only part of the full meson-baryon
scattering $T$-matrix. The $N^*$ excitations must be included. By extending
the $\pi N$ scattering formulation of Ref.~\cite{Sato:1996gk} to include
the $KY$ channel, one can show that the full meson-baryon scattering
amplitude can be written as
\begin{equation}
T_{\alpha,\beta}(E) = t_{\alpha,\beta}(E) + t^R_{\alpha,\beta}(E)\,,
\end{equation}
where $\alpha,\beta=\pi N, KY$. The nonresonant amplitudes
$t_{\alpha,\beta}$ are identical to those in
Eqs.~(\ref{eq:LS1})-(\ref{eq:LS2}). The resonant term at resonant energy
$E_{N^*}$ can be written in the familiar Breit-Wigner form (after
performing a proper diagonalization in the $N^*$ channel space)
\begin{equation}
t^R_{\alpha,\beta}(E)= \sum_{N^*}\frac{\bar{\Gamma}^*_{N^*,\alpha}\,
\bar{\Gamma}_{N^*,\beta}} {E - E_{N^*} +\frac{i}{2}\Gamma_{N^*}^{(tot)}}\,,
\end{equation}
with the total width
\begin{equation}
\Gamma_{N^*}^{(tot)} = \sum_{\alpha} |\bar{\Gamma}_{N^*,\alpha}|^2.
\end{equation}
In the above equations, $\bar{\Gamma}_{N^*,\alpha}$ describes the decay of
an $N^*$ resonance into a meson-baryon channel $\alpha$. Eqs.~(6)-(8) will
be the starting point for developing a strategy for using the empirical
$\pi N$ scattering amplitudes to solve Eqs.~(\ref{eq:LS1})-(\ref{eq:LS2}).

\section{Simplifications} \label{sec:Simp}
\begin{figure}
\centering
\renewcommand{\baselinestretch}{1.2}
\includegraphics[height=65mm]{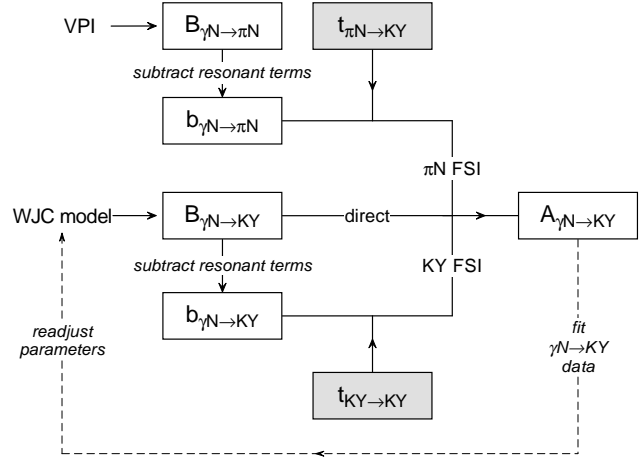}
\caption{Flow chart of our CC approach for kaon photoproduction.
 See the text for a description.}
\label{fig:CCchart}
\end{figure}

\begin{figure}
\centering
\renewcommand{\baselinestretch}{1.2}
\includegraphics[height=65mm]{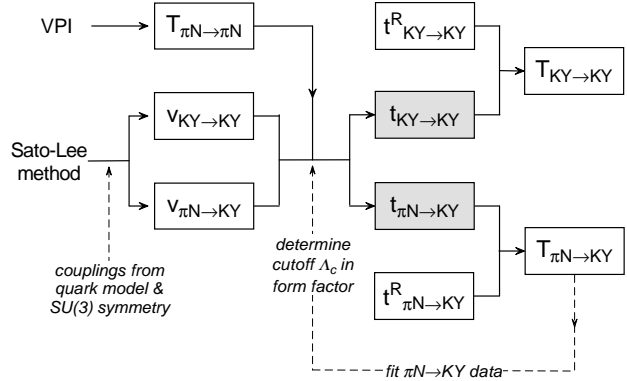}
\caption{Flow chart for the construction of the CC/FSI meson-baryon
$t$-matrices. See the text for a description.}
\label{fig:FSIchart}
\end{figure}

One way to approach the coupled-channel equations presented in the previous
section is to use effective Lagrangians to construct the photoproduction
operators $b_{KY,\gamma N}$ and $b_{\pi N,\gamma N}$, and meson-baryon
potentials $v_{\alpha,\beta}$. Then one could attempt to solve the full
$T$-matrix equations consistently as done in Ref.~\cite{Sato:1996gk} for
pion photoproduction. This program is too difficult and we therefore make
some simplifications in order to first gauge the role of the $\pi N$
channels in kaon photoproduction. The procedures used in this paper are
outlined in Figures~\ref{fig:CCchart} and \ref{fig:FSIchart}.


As indicated in Figure~\ref{fig:CCchart}, we take the isobar model
developed by Williams, Ji and Cotanch (WJC)~\cite{Williams:1992tp} to
describe the direct kaon photoproduction. The WJC model contains both the
resonant and nonresonant amplitudes, but does not include the meson-baryon
final state interactions. Thus, we can identify the amplitude generated
from this isobar model as $B_{KY,\gamma N}$ in Eq.~(\ref{eq:AB"}). The
nonresonant term $b_{KY,\gamma N}$ needed for evaluating Eq.~(\ref{eq:ab"})
can then be obtained from
\begin{equation}
b_{KY,\gamma N}= [B_{KY,\gamma N}]_{WJC} - [R_{KY,\gamma N}]_{WJC}
\end{equation}
where the resonant part $[R_{KY,\gamma N}]_{WJC}$ is also from the WJC
model. The ``subtract resonance term" procedure Eq.~(9) is indicated in the
lower left part of Fig.~\ref{fig:CCchart}.

Turning to the upper part of Figure~\ref{fig:CCchart}, we do not compute
the amplitude of the $\gamma N \rightarrow \pi N$ process---that is a
complicated CC problem by itself. Instead, we start with the VPI
partial-wave amplitudes for pion photoproduction~\cite{Arndt:1996ak}. We
then define the nonresonant part of pion photoproduction amplitude by
subtraction
\begin{equation}
b_{\pi N,\gamma N} = [B_{\pi N,\gamma N}]_{VPI} - R_{\pi N,\gamma N},
\end{equation}
where the resonant amplitude is calculated from
\begin{equation} \label{eq:RKY}
R_{\pi N,\gamma N}(E)= \sum_{N^*}\frac{\bar{\Gamma}^*_{N^*,\pi N}\,
\bar{\Gamma}_{N^*,\gamma N}} {E - E_{N^*} +\frac{i}{2}\Gamma_{N^*}^{(tot)}}.
\end{equation}
The partial decay widths $\bar{\Gamma}_{N^*,\gamma N}$ and
$\bar{\Gamma}_{N^*,\pi N}$ and the total widths $\Gamma^{(tot)}_{N^*}$ can
be calculated from the parameters listed by the Particle Data Group (PDG).

Eqs.~(9) and (10) only define the on-shell matrix elements of the
nonresonant photoproduction amplitudes. To evaluate the second and the
third FSI terms in Eq.~(\ref{eq:AB"}), we need to define their off-shell
behavior. For simplicity, we set
\begin{equation} \label{eq:off}
b_{\alpha,\gamma N}(q,k;E) = b_{\alpha,\gamma N}(q_0,k;E)
\cdot \frac{F(q)}{F(q_0)}\,,
\end{equation}
where $\alpha = KY$ and $\pi N$, $k$ and $q_0$ are the on-shell photon and
meson momenta fixed by the total energy $E$, $q$ is the desired off-shell
value, and $F(q)$ is a form factor to be defined later.

We now introduce a procedure to calculate the nonresonant
meson-baryon amplitudes $t_{KY,K'Y'}$ and $t_{KY,\pi N},$ which are shown as
shaded boxes in Fig.~\ref{fig:CCchart}. The procedure for obtaining these
transition amplitudes is outlined in Fig.~\ref{fig:FSIchart}. We again
start with the VPI amplitude. By using Eqs.~(6)-(8), the on-shell
nonresonant part of $\pi N$ amplitude $t_{\pi N,\pi N}$ is then defined by
\begin{equation}
t_{\pi N,\pi N} = [T_{\pi N,\pi N}]_{VPI} - t^R_{\pi N,\pi N},
\end{equation}
where the resonant term $t^R_{\pi N,\pi N}$, defined by Eq.~(7), can be
calculated using the resonant parameters listed by PDG. We then use the
same off-shell extrapolation defined by Eq.~(\ref{eq:off}) to define the
half-off-shell $\pi N$ t-matrix which is needed to evaluate the matrix
element of the second term of Eq.~(\ref{eq:LS2}).

\begin{figure}
\centering
\includegraphics[width=90mm]{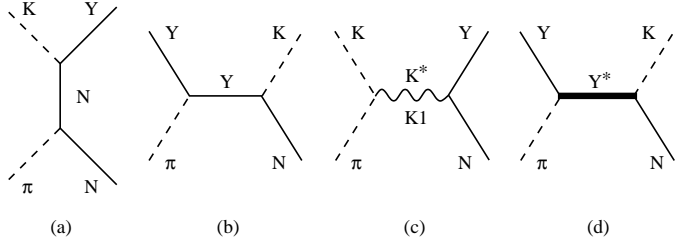}
\caption{Graphical representation of the potentials in
         $\pi N \rightarrow KY$.
         (a) direct nucleon pole $v_{N_D}$,
         (b) hyperon exchange $v_{Y_E}$,
         (c) strange vector meson exchange $v_{K^*}$, and
         (d) hyperon resonance exchange $v_{Y^*_E}$.}
\label{fig:piNKY}
\end{figure}

\begin{figure}
\centering
\includegraphics[width=90mm]{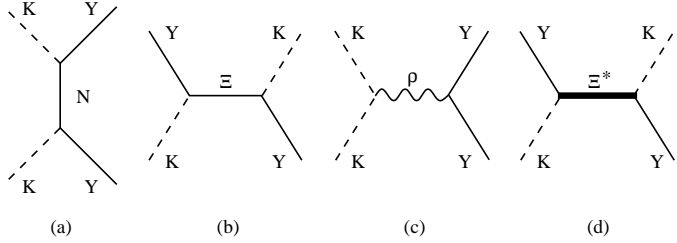}
\caption{Graphical representation of the potentials in
         $KY \rightarrow KY$.
         (a) direct nucleon pole $v_{N_D}$,
         (b) $\Xi$ exchange $v_{\Xi_E}$,
         (c) vector meson exchange $v_\rho$, and
         (d) $\Xi$ resonance exchange $v_{\Xi^*_E}$.}
\label{fig:KYKY}
\end{figure}

With the nonresonant $t_{\pi N,\pi N}$ defined by the above procedure,
Eqs.~(\ref{eq:LS1})-(\ref{eq:LS2}) can be solved by constructing the
potentials $v_{KY,KY}$ and $v_{KY,\pi N}$. Here we use the unitary
transformation method of Ref.~\cite{Sato:1996gk}. The considered potentials
are illustrated in Figs.~\ref{fig:piNKY} and \ref{fig:KYKY}. To solve the
coupled equations (\ref{eq:LS1})-(\ref{eq:LS2}), the meson-baryon
potentials must also be regularized by form factors. For simplicity, a form
factor $F(\mathbf{q}) = (\frac{\Lambda_c^2}{\Lambda_c^2+\mathbf{q}^2})^2$
is used to regularize all vertices in Figs.~\ref{fig:piNKY} and
\ref{fig:KYKY}, where $\mathbf{q}$ is the momentum of the external meson in
the center-of-mass frame.


To minimize the number of free parameters in this calculation, we fix most
of the couplings in Figs.~\ref{fig:piNKY} and \ref{fig:KYKY} by using
either the known PDG values~\cite{PDG:1998}, or from the predictions of
SU(3) flavor symmetry~\cite{Stoks:1996yj} or constituent quark
models~\cite{Capstick:1998uh,Koniuk:1980vy}. The coupling strengths of the
terms involving $\Xi$ and $\Xi^*$ are not known and therefore are not
included in this exploratory investigation. This of course should be
improved in later studies. To further simplify the calculation, the form
factor $F(q),$ used in defining the off-shell behavior of the nonresonant
photoproduction and $t_{\pi N,\pi N},$ is assumed to be the same as that
for regularizing the potentials $v_{KY,KY}$ and $v_{KY,\pi N}$. Thus, only
one cutoff $\Lambda_c$ needs to be determined.

\section{Results} \label{sec:Resul}

We start with the WJC model and hence the considered resonances and all
coupling strengths are fixed, as listed in the third column of Table 1. The
only parameter in our CC model is the cutoff $\Lambda_c$ of the form
factors which regularize $b_{\pi N,\gamma N}$, $b_{KY,\gamma N}$, $t_{\pi
N,\pi N}$, and potentials $v_{KY,KY}$ and $v_{KY,\pi N}$, as described in
Section~\ref{sec:Simp}. We determine this parameter by fitting the $\pi^- p
\rightarrow K^0\Lambda$ data. This fit is done by first solving
Eqs.~(\ref{eq:LS1})-(\ref{eq:LS2}) to obtain the nonresonant amplitude
$t_{K\Lambda,\pi N}$. The resonant part of this reaction is calculated from
using Eq.~(7) and the resonant parameters listed by PDG. We find that the
total cross section data of $\pi^- p \rightarrow K^0\Lambda$ can be fitted
by setting $\Lambda_c=680$ (MeV/c). This procedure then fixes the
nonresonant meson-baryon amplitudes $t_{KY,K'Y'}$ and $t_{KY,\pi N}$ that
are needed to evaluate the FSI effects on the photoproduction amplitude
using Eq.~(\ref{eq:AB"}).

\begin{figure}
  \centering
  \includegraphics[height=65mm]{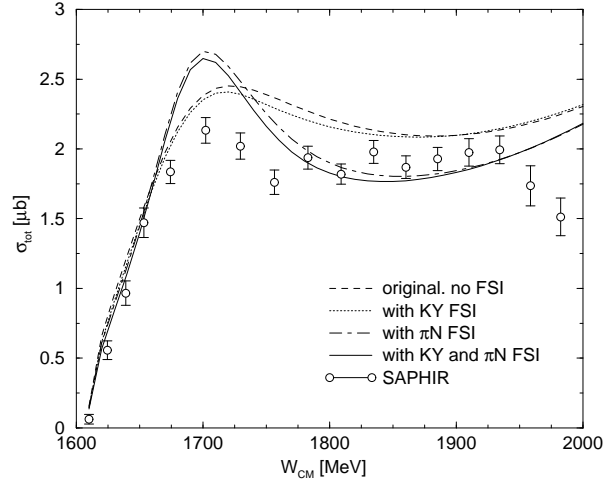}
  \caption{Total cross sections for $\gamma p\rightarrow K^+\Lambda$
           calculated using the CC approach
           with the original WJC coupling parameters.
           The curves are for the direct production $B_{K\Lambda}$
           (dashed line), $B_{K\Lambda}$ plus $K\Lambda$ FSI effects
           (dotted), $B_{K\Lambda}$ plus $\pi N$ FSI effects
           (dash-dotted), and full CC results $A_{K\Lambda}$ including
           $K\Lambda$ and $\pi N$ FSI (solid).
           Results are compared with the SAPHIR
           data~\cite{Tran:1998qw}.
           }
  \label{fig:gKtcs1}
\end{figure}

In Figure~\ref{fig:gKtcs1}, we illustrate the predicted FSI effects on the
total cross sections of $\gamma p\rightarrow K^+\Lambda$. Four curves are
shown: (1) the direct photoproduction calculated using the WJC model,
$B_{K\Lambda}$; (2) the direct production plus the $K\Lambda$ FSI effects,
$B_{K\Lambda} + t_{K\Lambda,K\Lambda}\, G_{K\Lambda}^{(+)}\, b_{K\Lambda}$;
(3) the direct production plus the $\pi N$ FSI effects, $B_{K\Lambda} +
t_{K\Lambda,\pi N}\,G_{\pi N}^{(+)}\,b_{\pi N}$; (4) the direct results
plus both the $K\Lambda$ and $\pi N$ FSI effects, i.e., our full CC results
$A_{K\Lambda}$. We also compare these results with the SAPHIR
data~\cite{Tran:1998qw}. Note that the WJC model was developed before the
SAPHIR measurement and hence their original fit (dashed curve) deviates
from these recent data.

\begin{figure}
  \centering
  \includegraphics[height=65mm]{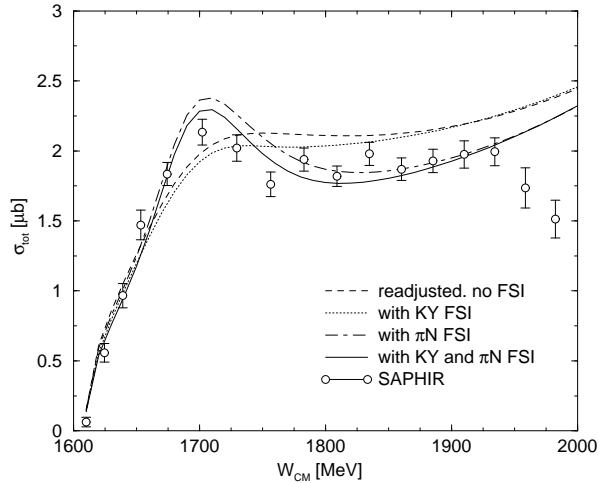}
  \caption{Total cross sections for $\gamma p\rightarrow K^+\Lambda$
           calculated using the CC approach
           with our readjusted coupling parameters.
           Curves and data as in Fig.~\ref{fig:gKtcs1}.
           }
  \label{fig:gKtcs2}
\end{figure}

In Figure~\ref{fig:gKtcs1}, we see significant differences up to 20\% of
the total cross sections with the $\pi N$ FSI (third term of
Eq.~(\ref{eq:AB"})) turned on and off. Clearly, the CC effects due to the
$\pi N$ channels make quite a sizable contribution to kaon photoproduction
and should be included in any kaon photoproduction calculation. Further
examination reveals that the major CC effects are from the s-wave $E_{0+}$
multipole. In contrast, the $K\Lambda$ FSI effect (second term of
Eq.~(\ref{eq:AB"})) is quite small.

Our full results (solid curve in Fig.~\ref{fig:gKtcs1}) based on the WJC
parameters deviate from the experimental data. We find that the data can
not be reproduced by only changing the original WJC parameters. Here we
emphasize that WJC published their predictions prior to the SAPHIR
measurements. In trying to fit this new data set, containing both
differential and total cross sections, we fail to reproduce the high energy
($W > 1950$ MeV) part. This failure is likely due to our lack of resonances
with mass around 1800 to 1900 MeV. It has been shown~\cite{saghai} that
good agreement with SAPHIR data is obtained if one introduces two spin-3/2
resonances $N^*$(1720) and $\Lambda ^*$(1890), provided off-shell effects
are included~\cite{Mizutani:1997sd}. However, including those two spin-3/2
resonances, with off-shell dynamics, requires 3 extra free parameters per
resonance. Our CC study aims to delineate the role of FSI and we wish to
keep the number of free parameters as small as possible. That is why we
have not yet introduced the requisite spin-3/2 resonances, but only
spin-1/2 resonances. One of these spin-1/2 resonances that we do include,
the $\Lambda ^*$(1810), is of some help in the relevant 1800 to 1900 MeV
mass range. We also note that a small contribution from the $N^*(1535)$
also improves the fit.

The extracted coupling constants (Table~\ref{tbl:coupl}, last column) show
that only two of the original WJC couplings get modified. The curve
corresponding to the parameter changes is depicted in
Fig.~\ref{fig:gKtcs2}, where we also show the results of the FSI decomposed
as in Fig.~\ref{fig:gKtcs1}. Comparison between Figs.~\ref{fig:gKtcs1} and
\ref{fig:gKtcs2} make clear that the numerical results for the FSI depend
on the resonance content of the reaction mechanism. However, this
dependence is very smooth in the case of the two configurations considered
here and does not alter the general trends nor the importance of the FSI
effects.

\begin{table}
\caption{Original fit parameters in the WJC model~\cite{Williams:1992tp}
and our readjusted values. The resonance couplings are the
products of photon and hadronic couplings.} \label{tbl:coupl}%
\centering
\smallskip
\begin{tabular}{cccc}
  \hline
  \ Particle\ & \ Coupling\ & \ WJC value\ & Readjusted value
  \\ \hline
  $\Lambda$ & $\frac{g_{K\Lambda N}}{\sqrt{4\pi}}$& -2.377 & -2.377
  \\[1ex] 
  $\Sigma^0$& $\frac{g_{K\Sigma N}}{\sqrt{4\pi}}$ &\ 0.222 &\ 0.404
  \\[1ex] 
  $K^*$     & $\frac{G^V_{K^*}}{\sqrt{4\pi}}$     & -0.162 & -0.162
  \\
  $  $      & $\frac{G^T_{K^*}}{4\pi}$            &\ 0.078 &\ 0.078
  \\[1ex] 
  $K1$      & $\frac{G^V_{K1}}{4\pi}$             &\ 0.019 &\ 0.019
  \\
  $  $      & $\frac{G^T_{K1}}{4\pi}$             &\ 0.173 &\ 0.173
  \\[1ex] 
  $N^*(1535)$      & $\frac{G_{N3}}{\sqrt{4\pi}}$        &  0     &\ 0.030
  \\[1ex] 
  $N^*(1650)$      & $\frac{G_{N4}}{\sqrt{4\pi}}$        & -0.044 & -0.025
  \\[1ex] 
  $N^*(1710)$      & $\frac{G_{N6}}{\sqrt{4\pi}}$        & -0.064 & -0.064
  \\[1ex] 
  $\Lambda^*(1405)$      & $\frac{G_{L1}}{\sqrt{4\pi}}$        & -0.073 & -0.073
  \\[1ex] 
  $\Lambda^*(1810)$      & $\frac{G_{L5}}{\sqrt{4\pi}}$        &  0     &\ 0.125
  \\[1ex] \hline
\end{tabular}
\end{table}

\section{Conclusions} \label{sec:Concl}
In this work we do not aim for an accurate reproduction of the data of kaon
photoproduction. We rather focus on the coupled-channel effects on this
reaction. The major conclusion from this study is that the $\pi N$ channels
make significant contributions through the coupled-channel mechanism and
must be included in a proper calculation for kaon photoproduction
reactions.

Our approach, based on an extension of the dynamical model of
Ref.~\cite{Sato:1996gk}, will be the basis for future investigations. In
particular, the $K\Sigma$ channels must be included in a more complete study
of kaon electromagnetic production, which is currently in progress. It is
especially interesting to study the $K\Sigma$ threshold effects (e.g.,
cusps at $K\Sigma$ threshold) in a CC calculation and their effect on spin
observables. In addition, more channels, such as $\eta N$, $\pi\Delta$ and
$\rho N$, must be included in a complete coupled-channel calculation.

\begin{ack}
The authors would like to thank S.A.~Dytman, R.~Schumacher, and S.N.~Yang
for helpful discussions. W.-T. C. is grateful to ANL and Saclay for the
hospitality extended to him during his visits. This work was supported in
parts by the U.S. National Science Foundation (PHY-9514885 and
PHY-9970775), the National Science Council of ROC under grant
No.~NSC89-2112-M002-078, a University of Pittsburgh Andrew Mellon
Predoctoral Fellowship, and U.S. DOE Nuclear Physics Division(Contract No.
W-31-109-ENG).
\end{ack}


\end{document}